\documentclass[reprint,amsmath,amssymb,aps,prx,superscriptaddress,longbibliography]{revtex4-1} 
\usepackage{amsmath}
\usepackage{amssymb}
\usepackage{mathtools}
\usepackage{physics}
\usepackage{graphicx}
\usepackage{float}
\usepackage[utf8]{inputenc}
\usepackage[colorlinks]{hyperref}

\newcommand\id{\mathbb{I}}
\newcommand\idq{\mathbb{I}_{2\times2}}

\begin{document}
    \title{Non-gaussian Entanglement Swapping between Three-Mode Spontaneous Parametric Down Conversion and Three Qubits}
    \author{A. Agustí Casado}
    \affiliation{Instituto de Física Fundamental, CSIC, Serrano 113-bis 28006 Madrid, Spain}
    \email{soyandres2@gmail.com}
    \author{C. Sabín}
    \affiliation{Departamento de Física Teórica, Univerdad Autónoma de Madrid E-28049 Madrid, Spain}
    \begin{abstract}
        In this work we study the production and swapping of non-gaussian multipartite entanglement in a setup containing a parametric amplifier which generates three photons in different modes coupled to three qubits. We prove that the entanglement generated in this setup is of nongaussian nature. We introduce witnesses of genuine tripartite nongaussian entanglement, valid both for mode and qubit entanglement. Moreover, those witnesses show that the entanglement generated among the photons can be swapped to the qubits, and indeed the qubits display nongaussian genuine tripartite entanglement over a wider parameter regime, suggesting that our setup could be a useful tool to extract entanglement generated in higher-order parametric amplification for quantum metrology or quantum computing applications.
    \end{abstract}
    \maketitle

    \section{Introduction}
        Entanglement is the key ingredient to most quantum technologies being designed today, ranging from teleportation \cite{tele, tele2, bouwmeester_pan1997} to boson sampling \cite{aaronson_arkhipov2013} and, in general, any quantum computational scheme. Therefore, plenty of present day literature deals with how to generate entanglement, and a very fruitful paradigm at that is parametric amplification. Take for example its role as a primitive ingredient in the recent claim on boson sampling quantum advantage \cite{zhong_wang2020}.

        The first instances of quantum parametric amplifiers date back to the 1980s \cite{slusher_hollberg1985, wu_kimble1986} in the setting of out-performing quantum measurements with single-mode squeezing. Then, in that same decade, it was discovered that parametric amplification could pump energy in two modes at once, leading to the generation of two-mode squeezing \cite{heidmann_horowicz1987}, perhaps the simplest form of continuous variable (CV) entanglement \cite{ou_pereira1992}. During the last five years, some of us have predicted that such two-mode squeezing can be used to entangle three modes in a genuinely tripartite way by applying the process to two pairs at once \cite{lahteenmaki_paraoanu2016,bruschi_sabin2017}, a prediction that has been experimentally validated \cite{chang_simoen2018}. We denominate this process double two-mode spontaneous parametric down-conversion (2-2SPDC). In a recent work \cite{agusti_chang2020}, we predicted that a similar process experimentally demonstrated in \cite{chang_sabin2020}, capable of generating three photons on different modes at once --three-mode spontaneous parametric down-conversion (3SPDC)-- produces genuine tripartite entanglement too. In order to experimentally detect 2-2SPDC entanglement, inspection of the covariances of the field quadratures was enough, whereas the 3SPDC entanglement requires inspecting higher statistical moments.

        As entanglement generation becomes a well stablished technology, produced in countless laboratories around the globe, still interesting theoretical questions remain open. Take, for example, the inequivalent entanglement of the three qubit \textit{W} and \textit{GHZ} states \cite{dur_vidal2000}. Those states are entangled in a tripartite way, and yet they can not be converted into each other by means of stochastic local operations and classical communication (SLOCC). A generalization of this result to general discrete-variables (DV) $d$-level systems has been recently proposed \cite{gharahi_mancini2021}, and the generalization to $n$ qubits is still incomplete, albeit we know that there have to be infinitely many SLOCC classes for $N > 3$ \cite{dur_vidal2000}, which therefore have to be gathered into some finite number of entanglement families --which proves to be a formidable task even for N=4  \cite{lamata1,lamata2,geometry1,geometry2}-- whose physical meaning is not always transparent. Furthermore, extensions of the above results to mixed states -even for three qubits- or to continuous variables (CV) beyond gaussian states remain as open problems. 
        A physically meaningful criterion to classify quantum entanglement, valid in principle both for CV and DV systems and for pure and mixed states, might be the distinction between gaussian and nongaussian entanglement. Besides the theoretical interest, nongaussian entanglement provides also technological advantages, for instance in quantum-metrology \cite{pesce, pesce2} or quantum computing applications \cite{laura}.

        In \cite{agusti_chang2020} we found that the states generated by 2-2SPDC and 3SPDC processes have different types of entanglement, suggesting some sort of continuous-variable equivalence with the three-qubit \textit{W} and \textit{GHZ} classes. In this work we formalize this insight, as well as analyze the swapping of entanglement from 3SPDC to three qubits. In particular, we provide formal definitions to gaussian and non-gaussian entanglement, and prove both the gaussianity of 2-2SPDC entanglement and the non-gaussianity of the  3SPDC entanglement, finding similarities and differences with GHZ and W classes.
        Moreover, we propose an experimental setup in which 3SPDC non-gaussian entanglement can be swapped to three qubits. An asymmetric SQUID generating 3SPDC is coupled to three separate resonators, each containing a coupled superconducting qubit. We show that the entanglement generated among the qubits is also of nongaussian nature, by using a natural extension of our CV entanglement witness which accommodates DV systems. Interestingly, we detect nongaussian qubit entanglement in a wider parameter regime -as compared to mode entanglement- which suggests that the swapping to qubits could be an efficient way of extending the technological usefulness of 3SPDC entanglement.

       The structure of this work will be as follows. In section II, we  introduce the notions of gaussian and non-gaussian entanglement in such a way that they may be applied to both CV and DV systems and pure and mixed states. Then, we relate these notions to the widely known \textit{W} and \textit{GHZ} states. After that, we present  arguments that can be used to prove the non-gaussianity of the entanglement contained in a state and we will apply them to our three-mode 3SPDC system in the presence of three qubits interacting each one with a bosonic mode. We will obtain proof of the tripartite non-gaussianity of the field's state, as well of the qubits'. Finally some concluding remarks and future research directions will be presented.

\section{Non-gaussian entanglement}
We start with a description of Non-gaussian entanglement.
        \label{section-non-gaussian-entanglement}
        \label{section-gaussian-entanglement}
        The term is coined after the gaussian states of quantum optics, those states represented by Wigner functions that happen to be gaussians of the canonical variables. 
Detecting entanglement in an experiment often involves measuring some witness, namely a combination of expectation values of observables that is bounded by some constant for states that do not posses the kind of entanglement considered.
        An entanglement witness is \textit{gaussian} if its algebraic expression contains only linear and quadratic contributions of the canonical variables. That way, the witness is only sensitive to the means and (co-)variances of a multipartite wave function or Wigner quasi-distribution. If higher powers of the canonical variables appear in the witness, or the witness can not be brought into an algebraic formula of the canonical variables, then it is \textit{non-gaussian}.
    
         The characterization of the entanglement of gaussian states is well known \cite{adesso2007}.  Any entanglement in a gaussian state will be detected by a gaussian witness -thus a gaussian state can only contain gaussian entanglement. However, a non-gaussian state might have the same mean and covariances of the canonical variables as some separable gaussian state \cite{agusti_chang2020}. Then, its entanglement would not be detected by a gaussian witness - and so it would be nongaussian entanglement.
     Finally, we can extend the concept of gaussianity to DV systems, by replacing any reference to canonical variables with spin variables.

         Interestingly, the concepts of gaussian and non-gaussian entanglement can be related with the two main representatives of tripartite qubit entanglement, the \textit{W} and \textit{GHZ} states. 
The \textit{W}-entanglement is gaussian, since can be detected by a gaussian witness \cite{teh_reid2019}, while
       \textit{GHZ}-entanglement is nongaussian, since we can for instance find a state that contains no entanglement and yet has the same means and covariances on the spin variables as the \textit{GHZ} state:
        \begin{align*}
            &\rho_{\text{mimic GHZ}} = \\
            &\frac{1}{12}
            \left(
                \ket{0_1}\bra{0_1} + \ket{1_1}\bra{1_1}
            \right)
            \otimes
            \left(
                \ket{0_20_3}\bra{0_20_3} + \ket{1_21_3}\bra{1_21_3}
            \right) \\
            &+
            \frac{1}{12}
            \left(
                \ket{0_2}\bra{0_2} + \ket{1_2}\bra{1_2}
            \right)
            \otimes
            \left(
                \ket{0_10_3}\bra{0_10_3} + \ket{1_11_3}\bra{1_11_3}
            \right) \\
            &+
            \frac{1}{12}
            \left(
                \ket{0_3}\bra{0_3} + \ket{1_3}\bra{1_3}
            \right)
            \otimes
            \left(
                \ket{0_10_2}\bra{0_10_2} + \ket{1_21_2}\bra{1_21_2}
            \right),
        \end{align*}
        where $\ket{0_i}$ is the ground state of the $i$-th qubit and $\ket{1_i}$ its excited state. Both the \textit{GHZ} state and the $\rho_{\text{mimic GHZ}}$ have the same first and second statistical moments of the spin variables
        \begin{alignat*}{3}
            \expval{S_x^i} &= 0 
            \quad \expval{S_y^i} &&= 0
            \quad \expval{S_z^i} &&= 0 \\
            \Delta^2S_x^iS_x^j &= 0 
            \quad \Delta^2S_y^iS_y^j &&= 0 
            \quad \Delta^2S_z^iS_z^j &&= 1/4
        \end{alignat*}
        where the spin variables are defined by $S_z^i\ket{0_i} = -1/2\ket{0_i}$ and $S_z^i\ket{1_i} = 1/2\ket{1_i}$ and the angular momentum algebra.
        

    \section{Non-gaussianity of entanglement in 3SPDC radiation}
       \label{section-3spdc+3qubits}

        The 3SPDC process studied in \cite{chang_sabin2020} takes place in a system composed of three bosonic modes subject to time-dependent boundary conditions, implemented by means of an asymmetric Superconducting Quantum Interference Device (SQUID), which behaves as a tunable non-linear inductor at the edge of a superconducting waveguide. The SQUIDs inductance is modulated with the sum of the characteristic frequencies of the three modes, producing an effective three-mode interaction described by
        \begin{align*}
            H_{\text{3SPDC-RWA}} = \sum_{i=1}^3 \omega_i a^\dagger_ia_i
            + g_0\cos\omega_dt
            \left(a^\dagger_1a^\dagger_2a^\dagger_3 + a_1a_2a_3\right)
        \end{align*}
        where $\omega_1, \omega_2$ and $\omega_3$ are the modes characteristic frequencies, $a^\dagger_i$ and $a_i$ the creation and annihilation operators on the $i$-th mode, $g_0$ the intensity of the coupling between the modes and $\omega_d$ is the driving to the SQUID, which is equal to $\sum_i\omega_i$. Note that the rotating wave approximation (RWA) was perfomed in order to illustrate the main process induced by this Hamiltonian: parametric creation or destruction of triplets of photons, one on each mode. The Hamiltonian is actually an approximation of a more general Hamiltonian
        \begin{align*}
            H_{\text{3SPDC}} &=
            \sum_{i=1}^3 \omega_i a^\dagger_ia_i \\
            &+ g_0\cos\omega_dt
            \left(a^\dagger_1+a_1\right)
            \left(a^\dagger_2+a_2\right)
            \left(a^\dagger_3+a_3\right),
        \end{align*}
        which will be the one that we will study throughout the text. We use this hamiltonian for the sake of completeness, although the RWA hamiltonian above would suffice to obtain the main results of this work and is generally valid under experimental conditions. However, using the general hamiltonian allows us not to worry with the regime of validity of the RWA.
         Before we begin proving the non-gaussian nature of the entanglement produced among the three modes, we will extend the system with three qubits, each one interacting with one mode. This modification is of interest because it paves the way to experimental production of non-gaussian entanglement both in CV systems (the reduced state of the three modes) and DV systems (the qubits). Such a technological platform could ground our theory on experimental data and, additionally, find technical applications as the primitive for generation of tripartite entanglement between CV or DV systems. 

\begin{figure}
            \centering
            \includegraphics[scale=1]{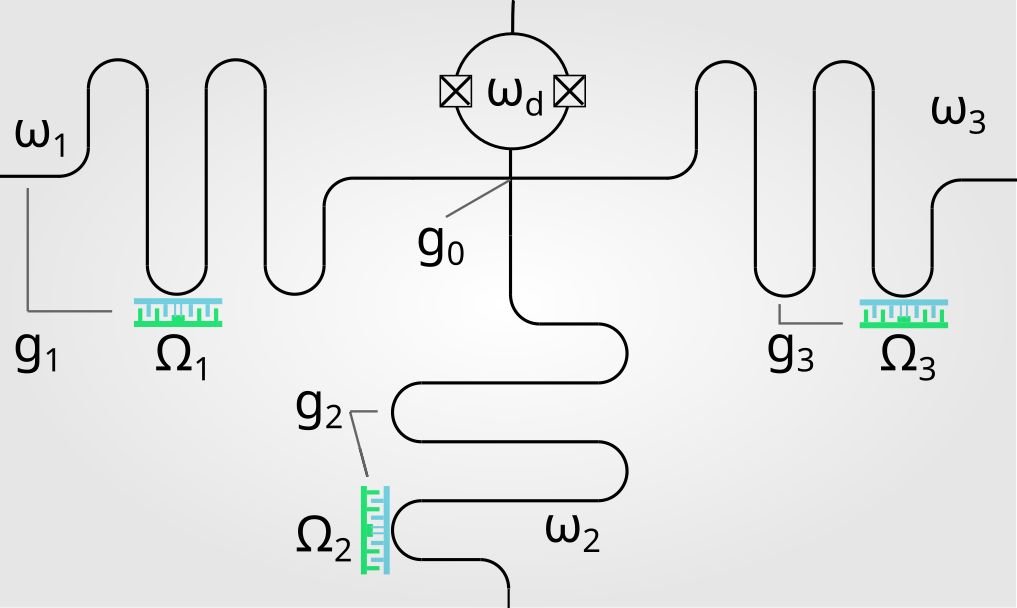}
            \caption{Illustration of the system composed of three transmission lines (depicted as solid meandered lines) that meet at an asymmetric SQUID (loop with boxes, that is, Josephson junctions, at the sides). Each one of those transmission lines interacts with a transmon qubit (colored zipers, not to scale). Control lines have been omitted. If pumped with the appropiate tone, the asymmetric SQUID will drive three-mode spontaneous parametric down conversion among the three fundamental modes of the transmission lines. Nongaussian tripartite entanglement will be produced between the modes as proved in Eq. (\ref{eq:constant-covariances}) and Fig. (\ref{sim-Gm.png}) for some parameter regimes. Additionally, nongaussian tripartite entanglement will be swapped to the qubits, as proved in the text and Fig. (\ref{sim-Gq.png}). We show labels for the parameters that appear in Hamiltonian Eq. (\ref{H}) for reference.}
            \label{fig:system}
        \end{figure}

        When the three qubits are taken into account, the total Hamiltonian becomes
        \begin{align}
            H_{\text{3SPDC+3qubits}} &=
            \sum_{i=1}^3 \omega_i a^\dagger_ia_i 
            + \frac{\Omega_i}{2}\sigma_{z, i}
            + g_i \sigma_{x, i}\left(a^\dagger_i + a_i\right) \nonumber\\
            &+ g_0\cos\omega_dt
            \left(a^\dagger_1+a_1\right)
            \left(a^\dagger_2+a_2\right)
            \left(a^\dagger_3+a_3\right),
            \label{H}
        \end{align}
        where $\sigma_{x, y, z, i}$ are the Pauli matrices for the $i$-th qubit and $g_i$ the intensity of its coupling to the respective mode. Note that the qubit-mode interaction takes the form of the Rabi interaction. An experimental setup that could be effectively modeled with Eq. (\ref{H}) is described in Figure (\ref{fig:system}). It is composed of three superconducting cavities joined together from one of their edges \cite{houck1,houck2012chip,nuestrosimone}. At that meeting point lies an asymmetric SQUID driven with a single tone of frequency $\omega_d = \sum_i\omega_i$. 

        In order to prove the non-gaussianity of the entanglement produced by Hamiltonian in Eq. (\ref{H}) when evolving the initial vacuum state $\ket{0g0g0g}$, where $\ket{0}$ is the mode vacuum state and $\ket{g}$ is the qubit ground state, we will examine the time derivatives of the quadratures and spin covariances, by making use of the following condition
        \begin{align}
          i\hbar\partial_t\Delta^2&O_iO_j = 0 \nonumber \\
            &\Leftrightarrow \nonumber \\
           \expval{[O_iO_j,H]}
            &= \expval{O_i}\expval{[O_j, H]}
            + \expval{[O_i, H]}\expval{O_j}
            \label{constant-covariance-condition}
       \end{align}
        where $O_i$ and $O_j$ are canonical or spin variables, $H$ is the Hamiltonian of the system and $\Delta^2O_iO_j$ is the covariance between the measurements of $O_i$ and $O_j$, that is $\expval{O_iO_j} - \expval{O_i}\expval{O_j}$. Eq. (\ref{constant-covariance-condition}) is easily derived from the Heisenberg equation of motion. See Appendix \ref{section-constant-covariance-condition} for further notes on its derivation.
        Using the Hamiltonian in Eq. (\ref{H}) and Eq. (\ref{constant-covariance-condition}), we have:
        \begin{alignat}{1}
            \partial_t \Delta^2 x_ix_j &= 
            \expval{\frac{x_ip_j}{m_j} 
            + \frac{x_jp_i}{m_i}}
            - \expval{x_i}\expval{\frac{p_j}{m_j}}
            - \expval{\frac{p_i}{m_i}}\expval{x_j}.
            \nonumber \\
            \partial_t \Delta^2 p_ip_j &= 
            -\expval{
            m_j\omega_j^2p_ix_j 
            + m_i\omega_i^2x_ip_j
            } \nonumber \\
            &- i\hbar\expval{g_j\sigma_{xj}p_i + g_i\sigma_{xi}p_j}
            \nonumber \\
            &- \hat{g}(t)\expval{
            p_ix_ix_k + x_jp_jx_k} \nonumber \\
            &+ \expval{m_i\omega_i^2 x_i
            + g_i\sigma_{xi}
            + \hat{g}(t)x_jx_k}\expval{p_j} \nonumber \\
            &+ \expval{m_j\omega_j^2 x_j
            + g_j\sigma_{xj}
            + \hat{g}(t)x_ix_k}\expval{p_i}
            \nonumber \\
            \partial_t \Delta^2 S_{xi}S_{xj} &= 
            \Omega_i\expval{\sigma_x^i\sigma_y^j}
            + \Omega_j \expval{\sigma_y^i\sigma_x^j}
            \nonumber \\
            \partial_t \Delta^2 S_{yi}S_{yj} &= 
            \Omega_j \expval{\sigma_y^i\sigma_x^j}
            + \Omega_i\expval{\sigma_x^i\sigma_y^j}
            \nonumber \\
            \partial_t \Delta^2 S_{zi}S_{zj} &= 
            \frac{g_j}{2}\expval{\sigma_z^ix_j\sigma_y^j}
            + \frac{g_i}{2}\expval{x_i\sigma_y^i\sigma_z^j}
            \label{covariance-time-derivative}
        \end{alignat}
        where $x_i$ and $p_i$ are the quadratures of the $i$-th mode and $S_{x, y, z, i}$ are the \textit{analog angular momentum} operators along the $x$, $y$ and $z$ axes for the $i$-th qubit. For a detailed derivation of the covariances time derivatives see Appendix \ref{appendix-covariance-time-derivative}. In order to tackle Eqs. (\ref{covariance-time-derivative}) we consider the following projector
        \begin{align}
            P &= 
            \sum_{\alpha, \beta=0}^1
            \bigotimes_{i=1}^3\sum_{n=0}^\infty
            P_i(2n+\alpha) \otimes P_{i,2\times2}(\beta)
            \label{conserved-quantity}
        \end{align}
        where $P_i(n) = \ketbra{n}$ is the projector onto the bosonic mode state with $n$ photons or excitations and $P_{i,2\times2}(q)$ is $\ketbra{g}$ if $q=0$, the projector onto the qubit ground state, or $\ketbra{e}$ if $q=1$, the projector onto the qubit excited state. We find that this projector is a conserved quantity of the system. Please consider the following motivation behind its definition: the Hamiltonian in Eq.(\ref{H}) allows for some transitions between the stationary Hamiltonian eigenstates. In particular, it allows for transitions that change all three modes in one photon (via the 3SPDC process) as well as transitions changing a qubit-mode pair in one excitation (that is, any combination of creating or destroying a photon while exciting or relaxing the qubit). But there are many other transitions that are not allowed: creating/destroying a pair of photons but not a third one, spontaneously exciting or relaxing a qubit without changing photon number, and so on. Then, P is built to project onto all of the eigenstates the vacuum can transition to, while excluding those the vacuum can not leak into. For further information about the derivation of P, as well as proof of how it commutes with the Hamiltonian, see Appendix \ref{appendix-conserved-quantity}. The expectation value of $P$ for the initial state $\ket{0g0g0g}$ is 1. Therefore, the time evolution of $\ket{0g0g0g}$ will never leave the subspace $P$ projects onto, which we denote \textit{the dynamical subspace} 
        \begin{align*}
            \psi(t) = 
            \sum_{\alpha, \beta=0}^1
            \bigotimes_{i=1}^3
            \sum_{n=0}^\infty
            c_{\alpha, \beta, i, n}(t)\ket{2n+\alpha}\otimes\ket{\beta}
        \end{align*}
        With this we can evaluate many of the expectation values in the covariances time derivatives in Eq. (\ref{covariance-time-derivative}). In particular, all time derivatives become zero, except for the $\Delta^2S_{z, i}S_{z, j}$ covariance
        \begin{align}
            \partial_t \Delta^2 x_ix_j &= 0
            \nonumber \nonumber \\
            \partial_t \Delta^2 p_ip_j &= 0
            \nonumber \nonumber \\
            \partial_t \Delta^2 S_{xi}S_{xj} &= 0
            \nonumber \nonumber \\
            \partial_t \Delta^2 S_{yi}S_{yj} &= 0
            \nonumber \nonumber \\
            \partial_t \Delta^2 S_z^i S_z^j &=
            \frac{g_j}{2}\expval{\sigma_z^ix_j\sigma_y^j}
            + \frac{g_i}{2}\expval{x_i\sigma_y^i\sigma_z^j} \neq 0
            \label{eq:constant-covariances}
        \end{align}
       
\begin{figure}
            \centering
            \includegraphics[scale=.3]{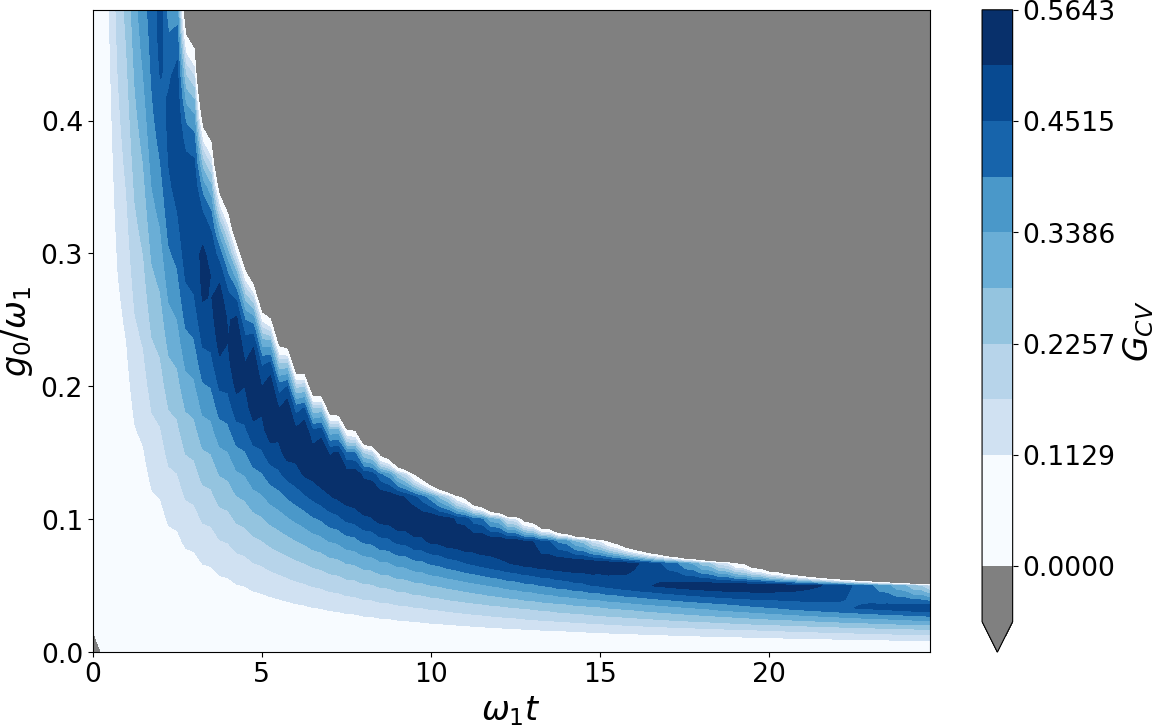}
            \caption{Value of the witness $G_{CV}$ defined in Eq. (\ref{Gm}) as a function of time $t$ and 3SPDC coupling $g_0$ in units of the lowest frequency mode $\omega_1$ when the initial state $\ket{0g0g0g}$ evolves under the Hamiltonian in Eq. (\ref{H}). The other mode frequencies are $\omega_2 = 2\omega_1$ and $\omega_3 = \omega_1$. The qubits are resonant with their modes so that $\Omega_i = \omega_i$ and their couplings are all equal $g_i = 0.01\omega_1$. The witness reports non-gaussian entanglement in the modes, that is, it is greater than zero, for short times. Please note that entanglement is 0 at $t=0$.}
            \label{sim-Gm.png}
        \end{figure}

        Therefore, the reduced state of the three modes can not contain gaussian entanglement: it has the same covariances than a clearly separable state, the vacuum $\ket{000}$. But the state gets entangled with time, as we proved in \cite{agusti_chang2020} for the qubit-less system. In that work we built a genuine tripartite entanglement witness defined
        \begin{align*}
            G'_{CV} = \abs{\expval{a_1a_2a_3}} 
            - \sum_{\substack{i, j, k = 1, 2, 3 \\ i \neq j \neq k \neq i}}
            \sqrt{
                \expval{a^\dagger_ia_i}
                \expval{a^\dagger_ja_ja^\dagger_ka_k}
            }
        \end{align*}
        so that when $G'_{CV} > 0$ genuine tripatite entanglement is detected. In fact, since the publication of \cite{agusti_chang2020} we have found an improved witness
        \begin{equation}
            G_{CV} = \abs{\expval{a_1a_2a_3}} 
            - \max_{\substack{i, j, k = 1, 2, 3 \\ i \neq j \neq k \neq i}}
            \sqrt{
                \expval{a^\dagger_ia_i}
                \expval{a^\dagger_ja_ja^\dagger_ka_k}
            }
            \label{Gm}
        \end{equation}
        by following the derivation in \cite{agusti_chang2020} and making use of the fact that the expectation values of a mixed state cannot be larger than the largest of its pure components. Figure (\ref{sim-Gm.png}) shows the value of the genuine tripartite entanglement witness $G_{CV}$ for different times and 3SPDC coupling strength. We conclude that the field contains non-gaussian entanglement at times not much larger than $g_0\,t =1$. For larger times, all we know is that gaussian witnesses will fail, but if there is any entanglement in the modes non-gaussian witnesses might succeed.
\section{Non-gaussian three-qubit entanglement}
        The nature of the three-qubit entanglement is, however, more difficult to determine: since the $z$ covariances do change in time we need to answer the question of whether or not a gaussian witness exists that uses only the $z$ spin covariances. We find that the answer is no, and therefore the qubit entanglement, if there is any, is non-gaussian too. See Appendix \ref{appendix-z-spin-covariance} for a proof. 
        
        In order to detect whether there is actually entanglement, we need a suitable nongaussian entanglement witness. The same proof  \cite{agusti_chang2020} that lead to the construction of $G_{CV}$ in CV systems can be extended to a DV witness by replacing the canonical variables with spin variables
        \begin{equation}
            G_{DV} = \abs{\expval{\sigma^-_1\sigma^-_2\sigma^-_3}} 
            - \max_{\substack{i, j, k = 1, 2, 3 \\ i \neq j \neq k \neq i}}
            \sqrt{
                \expval{\sigma^+_i\sigma^-_i}
                \expval{\sigma^+_j\sigma^-_j\sigma^+_k\sigma^-_k}
            }
            \label{Gq}
        \end{equation}
        \begin{figure}[b]
            \centering
            \includegraphics[scale=.3]{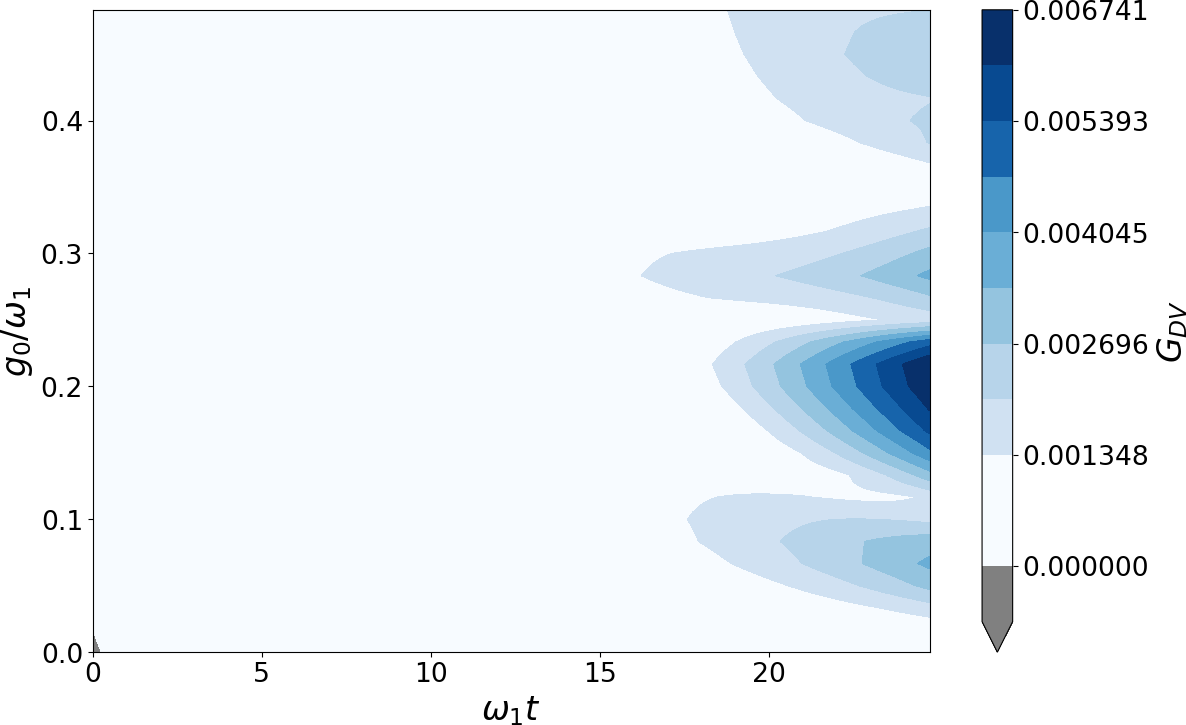}
            \caption{Value of the witness $G_{DV}$ defined in Eq. (\ref{Gq}) as a function of time $t$ and 3SPDC coupling $g_0$ in units of the lowest frequency mode $\omega_1$ in the same conditions as Figure (\ref{sim-Gm.png}). The witness reports non-gaussian entanglement in the qubits, that is, it is greater than zero, for a broad parameter regime. Please note that entanglement is 0 at $t=0$.}
            \label{sim-Gq.png}
        \end{figure}
        which works as $G_{CV}$ but in DV systems, it reports genuine tripartite entanglement whenever $G_{DV} > 0$. Figure (\ref{sim-Gq.png}) shows the value of $G_{DV}$ for different times and 3SPDC coupling strengths. We conclude that the qubits are, indeed, entangled in a non-gaussian way for a broad parameter regime. Indeed, it seems that the qubits are entangled in a wider regime of parameters, suggesting that swapping the entanglement from the photons to the qubits could be a way to exploit the multipartite entanglement generated in 3SPDC radiation. However, notice that there could be other witnesses detecting entanglement where ours fails. Note also that, as usual, an entanglement witness only tells us about the existence of entanglement, not necessarily its degree, which would require the use of an entanglement measure.

    \section{Conclusions and Future directions}
    In summary, we have presented a setup in which three qubits are coupled to a 3SPDC source. We have shown that there is genuine tripartite entanglement generated both among the three modes of the electromagnetic field and among the qubits. Moreover, we have proved the nongaussian nature of this entanglement, as well as that of the \textit{GHZ} state, suggesting that gaussianity might be an extension to CV and mixed states of the \textit{W} and \textit{GHZ} classes. We have introduced witnesses of genuine tripartite entanglement both for the field and the qubits. Interestingly, in the case of the qubits, entanglement is detected for a wider regime of parameters, which suggests that our setup could provide an efficient way of exploiting the genuine nongaussian multipartite entanglement generated in 3SPDC interactions. In particular, qubits with nongaussian entanglement display useful properties for quantum-metrology and quantum-computing applications.

\section*{Acknowledgements}

 A.A.C acknowledges support from Postdoctoral Junior Leader Fellowship Programme from la Caixa Banking Foundation (LCF/BQ/LR18/11640005). C. S acknowledges support from Spanish Ramón y Cajal Program RYC2019-028014-I.

    \appendix

    \section{Dynamics of statistical moments}
        \label{section-constant-covariance-condition}
        In this Appendix we will derive the expression for the time derivatives of the canonical and spin variables covariances. We will be particularly interested in the cases when the moments are constant. If that is the case, gaussian entanglement can not be generated. We start with the Heisenberg equation of motion
        \begin{equation}
            i\hbar\partial_t O(t) = [O(t),H(t)],
        \end{equation}
        which immediately yields expressions for the time derivatives of the first order statistical moments, the means
        \begin{align}
            i\hbar\partial_t\expval{O} = \expval{[O(t), H(t)]}
        \end{align}
        In order to derive a similar expression for second order statistical moments, that is, variances and covariances, we follow a similar approach. We recall the definition of the covariances of two observables $O_i$ and $O_j$
        \begin{align*}
            \Delta^2O_iO_j = \expval{O_iO_j} - \expval{O_i}\expval{O_j}
        \end{align*}
        and by taking its time derivative one arrives at
        \begin{align*}
            i\hbar\partial_t\Delta^2O_iO_j
            = \expval{[O_iO_j, H]} 
            - \expval{O_i}\expval{[O_j, H]}
            - \expval{O_j}\expval{[O_i, H]}
        \end{align*}
        This equation gives us conditions systems must follow in order not to generate or destroy gaussian entanglement
        \begin{align}
            i\hbar\partial_t\Delta^2&O_iO_j = 0 \nonumber \\
            &\Leftrightarrow \nonumber \\
            \expval{[O_iO_j,H]}
            &= \expval{O_i}\expval{[O_j, H]}
            + \expval{[O_i, H]}\expval{O_j}
        \end{align}
        Note that if the averages of $O_i, O_j$ are zero, then the condition states that in order not to change the covariances, the operator $O_iO_j$ must be a conserved quantity in the subspace spanned by the state during all that time.

        Summarizing, we have obtained expressions for the time derivatives of the means and covariances of general observables. Those equations have lead to Hamiltonian conditions in Eq. (\ref{constant-covariance-condition}) that will tell when the covariances (and gaussian entanglement) are constant in a particular system. 
        We will consider particular Hamiltonians in the calculations to come.
    
    \section{Derivation of the covariances' time-derivatives}
        \label{appendix-covariance-time-derivative}
        In this Appendix we will take Hamiltonian in Eq. (\ref{H}) and compute the covariances' time-derivatives as instructed by Eq. (\ref{constant-covariance-condition}). Note that the Hamiltonian can be written in terms of the canonical and spin variables alone
        \begin{align*}
            H &=
            \sum_{i=1}^3
            \left[
            \frac{p_i^2}{2m_i} 
            + \frac{1}{2}m_i\omega_i^2x_i^2
            + \Omega S_z^i
            \right] \nonumber \\
            &+ \hat{g}_0\cos(\sum_i\omega_it)x_1x_2x_3
            + \sum_{i=1}^3 g_i \sigma_x^ix_i
        \end{align*}
        Then, the field's position covariances have the following time derivatives
        \begin{align}
            [x_i, H] 
            &= \frac{1}{2m_i}[x_i, p_i^2] \nonumber \\
            &= \frac{1}{2m_i}([x_i, p_i]p_i + p_i[x_i, p_i]) \nonumber \\
            &= \frac{i\hbar}{m_i}p_i \nonumber \\
            [x_ix_j, H]
            &= x_i[x_j, H] + [x_i, H]x_j \nonumber \\
            &= i\hbar\left(\frac{x_ip_j}{m_j} + \frac{x_jp_i}{m_i}\right) \nonumber \\
            \partial_t \Delta^2 x_ix_j
            &= \expval{\frac{x_ip_j}{m_j} 
            + \frac{x_jp_i}{m_i}}
            - \expval{x_i}\expval{\frac{p_j}{m_j}}
            - \expval{\frac{p_i}{m_i}}\expval{x_j}.
            \label{position-covariance-time-derivative-norwa}
        \end{align}
        And for the momentum's covariances
        \begin{align*}
            [p_i, H] 
            &= \frac{m_i\omega_i^2}{2}[p_i, x_i^2] 
            + g_i\sigma_{xi}[p_i, x_i]
            + \hat{g}(t)[p_i, x_1x_2x_3] \\
            &= - i\hbar m_i\omega_i^2 x_i 
            - i\hbar g_i\sigma_{xi}
            - i\hbar\hat{g}(t)x_jx_k  \\
            & \text{ with } i \neq j \neq k \neq i \\
            [p_ip_j, H] 
            &=  p_i[p_j,H] 
            + [p_i, H]p_j \\
            &= -i\hbar\left(
            m_j\omega_j^2p_ix_j 
            + m_i\omega_i^2x_ip_j
            \right) \\
            &- i\hbar\left(g_j\sigma_{xj}p_i + g_i\sigma_{xi}p_j\right)\\
            &- i\hbar\hat{g}(t)\left(
                p_ix_ix_k + x_jp_jx_k
            \right)
        \end{align*}
        which results in a time derivative of the momenta covariances
        \begin{align}
            \partial_t \Delta^2p_ip_j
            = &-\expval{
            m_j\omega_j^2p_ix_j 
            + m_i\omega_i^2x_ip_j
            } \nonumber \\
            &- i\hbar\expval{g_j\sigma_{xj}p_i + g_i\sigma_{xi}p_j}
            \nonumber \\
            &- \hat{g}(t)\expval{
            p_ix_ix_k + x_jp_jx_k} \nonumber \\
            &+ \expval{m_i\omega_i^2 x_i
            + g_i\sigma_{xi}
            + \hat{g}(t)x_jx_k}\expval{p_j} \nonumber \\
            &+ \expval{m_j\omega_j^2 x_j
            + g_j\sigma_{xj}
            + \hat{g}(t)x_ix_k}\expval{p_i}
            \label{momentum-covariance-time-derivative-norwa}
        \end{align}
        The conditions derived in Eq. (\ref{constant-covariance-condition}) not only apply to continuous variables systems, but discrete ones as well. By plugging the spin variables $S_x^i$, $S_y^i$ and $S_z^i$ as well as the Hamiltonian in Eq. (\ref{H}) we derive
        \begin{align*}
            \partial_t \Delta^2 S_x^i S_x^j &=
            \Omega_i\expval{\sigma_x^i\sigma_y^j}
            + \Omega_j \expval{\sigma_y^i\sigma_x^j} \\
            \partial_t \Delta^2 S_y^i S_y^j &=
            \Omega_j \expval{\sigma_y^i\sigma_x^j}
            + \Omega_i\expval{\sigma_x^i\sigma_y^j} \\
            &- 2g_j\expval{x_j\sigma_y^i\sigma_z^j}
            - 2g_i\expval{x_i\sigma_z^i\sigma_y^j} \\
            \partial_t \Delta^2 S_z^i S_z^j &=
            \frac{g_j}{2}\expval{\sigma_z^ix_j\sigma_y^j}
            + \frac{g_i}{2}\expval{x_i\sigma_y^i\sigma_z^j}
        \end{align*}

    \section{Conserved quantities}
        \label{appendix-conserved-quantity}
        In this appendix we will provide proof of the conserved quantity \textit{P} in Eq. (\ref{conserved-quantity}). Note that \textit{P} projects onto the subspace that contains every eigenstate with the same parity of qubit plus photon excitation on each pair of qubits and modes. That is, for every eigenstate in that subspace, the addition of the number of photons on the first mode plus the number of excitations on the first qubit (that is, zero for $\ket{g}$ or one for $\ket{e}$) will always be the same that the addition of the number of photons and qubit excitations in the second qubit-mode pair. The same happens with the third qubit-mode pair. In order to gain some insight on why that particular projector is a conserved quantity we will first argue for its construction with perturbation theory. Then, an actual proof calculating the commutator with the Hamiltonian is provided. Finally, we will compute some elementary expectation values within the image of \textit{P} that happen to appear in the covariances' time-derivatives.
        \subsection{Construction of a conserved quantity}
            \label{section-3SPDC+3qubits-P-construction}
            We will begin with the first order perturbative corrections to the time evolution of $H_{\text{3SPDC+3qubits}}$
            \begin{align*}
                \psi^{(1)}(t) &= \frac{1}{i\hbar}\int_0^tdt'
                H_{\text{int}}(t')\ket{000ggg} \\
                &= \alpha\ket{111ggg} 
                + \beta\ket{100egg} 
                + \gamma\ket{010geg} 
                + \delta\ket{001gge} 
            \end{align*}
            where $H_{\text{int}}$ is the Hamiltonian in the interaction picture. The important fact to note here is that all kets share some short of parity. If we add together the number of photons in the first mode and the number of excitations in the first qubit we obtain 2 or 0, even numbers. The same happens with every pair mode-qubit and for every ket.

            The second order correction takes the form
            \begin{alignat*}{2}
                \psi^{(2)}(t) && = \frac{1}{i\hbar}\int&_0^tdt'
                H_{\text{int}}(t')\psi^{(1)}(t') \\
                &&\in \text{span}(&
                \ket{000ggg},
                \ket{002ggg},
                \ket{020ggg},
                \ket{022ggg}, \\
                && &
                \ket{200ggg},
                \ket{202ggg},
                \ket{220ggg},
                \ket{222ggg}, \\
                && &
                \ket{110gge},
                \ket{101geg},
                \ket{011egg},
                \ket{112gge}, \\
                && &
                \ket{121geg},
                \ket{211egg},
                \ket{211egg},
                \ket{011egg}, \\
                && &
                \ket{000ggg},
                \ket{200ggg},
                \ket{110eeg},
                \ket{101ege}, \\
                && &
                \ket{121geg},
                \ket{101geg},
                \ket{110eeg},
                \ket{000ggg}, \\
                && &
                \ket{020ggg},
                \ket{011gee},
                \ket{112gge},
                \ket{110gge}, \\
                && &
                \ket{101ege},
                \ket{011gee},
                \ket{002ggg},
                \ket{000ggg}
                ),
            \end{alignat*}
            again, all the kets involved in the second order correction share a notion of parity. But it appears to be a different, or more general, parity than the first order corrections. Some kets have an even number of photons plus qubit excitations (e.g. $\ket{222ggg}$). Other kets have an odd number of photons plus qubits excitations (e.g. $\ket{110gge}$). But there are no kets that mix odd and even numbers of photons plus qubit excitations (e.g. there is no $\ket{211geg}$).
                        
            The reader might have noticed that we are now in position to finish a proof by induction. We have proven that the first order corrections are composed of kets with even number of field plus qubit excitations. We have proven that the second order corrections are a superposition of kets with odd \textit{or} even (but no mixtures) of field plus qubit excitations. Now we will prove that if the $n$-th order correction is such a superposition, the $n+1$-th correction has that same parity. In order to do so, we will study the effects each of the pieces of the Hamiltonian have on the parity of a ket.
                        
            Firstly, the 3SPDC piece. It has the form $g(t)(a^\dagger_1 + a_1)(a^\dagger_2 + a_2)(a^\dagger_3 + a_3)$. Note that the result of the application of this piece of the Hamiltonian on a vector with well defined parity is to completely change the parity of each mode-qubit pair. That is, each mode has to change its number of photons in one unit, up or down, but their interacting qubit will remain the same. Therefore, the result is a superposition of vectors with the same parity on each qubit-mode pair.
            
            Secondly, the Rabi piece. If has the form $g_i(t)\sigma_x^i(a^\dagger_i + a_i)$. The result of applying this piece of the Hamiltonian on a vector with well defined parity is a superposition of vectors of the same parity. This is due to the fact that the $i$-th qubit must change its quantum number and the same $i$-th mode must change its number of photons in one unit. Therefore the parity of that pair will be the same.
                
            Because of these two facts, the parities of the kets forming the superposition that is the evolution of vacuum will never mix. And therefore, the state must remain in the subspace of vectors with well defined qubit plus mode excitation parity. The operator that projects onto the subspace of vectors with that well defined excitation parity is
            \begin{align}
                P &= 
                \bigotimes_{i=1}^3\sum_{n=0}^\infty
                P_i(2n) \otimes P_{i,2\times2}(0) \nonumber \\
                &+
                \bigotimes_{i=1}^3\sum_{n=0}^\infty
                P_i(2n+1) \otimes P_{i,2\times2}(0) \nonumber \\
                &+
                \bigotimes_{i=1}^3\sum_{n=0}^\infty
                P_i(2n) \otimes  P_{i,2\times2}(1) \nonumber \\
                &+
                \bigotimes_{i=1}^3\sum_{n=0}^\infty
                P_i(2n+1) \otimes P_{i,2\times2}(1) \nonumber \\
                &= \sum_{\alpha, \beta=0}^1
                \bigotimes_{i=1}^3\sum_{n=0}^\infty
                P_i(2n+\alpha) \otimes P_{i,2\times2}(\beta)
            \end{align}
            where $P_i(n)$ is the Fock state projector $\ket{n}\bra{n}$ and $P_{i,2\times2}(q)$ is the projector onto the $S_z$ lower eigenstate if $q = 0$ or onto the higher eigenstate if $q = 1$.
    
        \subsection{Proof that $P$ is a conserved quantity}
        
          In this section, we let the i indices drop as they are redundant notation  The projector $P$ clearly commutes with the Hamiltonian's stationary part. In order to prove that it commutes with the interacting pieces as well we need to introduce some notation
            \begin{align*}
                x_1x_2x_3 &\rightarrow
                \bigotimes_{i=1}^3 x \otimes \mathbb{I}_{2\times2} \\
                \sigma_x x &\rightarrow
                \bigotimes_{j=1}^3(
                    \delta_{ij}x\otimes\sigma_x
                    + (1 - \delta_{ij})
                    \mathbb{I}\otimes\mathbb{I}_{2\times2})
            \end{align*}
            First, we will show that $x_1x_2x_3$ commutes with $P$
            \begin{widetext}
            \begin{alignat*}{2}
                x_1x_2x_3P &=
                \bigotimes_{i=1}^3 x\otimes&&\idq
                \sum_{\alpha, \beta=0}^1
                \bigotimes_{i=1}^3
                \sum_{n=0}^\infty
                P(2n + \alpha) \otimes P_{2\times2}(\beta) \\
                &= 
                \sum_{\alpha, \beta=0}^1
                \bigotimes_{i=1}^3&&
                \sum_{n=0}^\infty
                xP(2n + \alpha) \otimes P_{2\times2}(\beta) \\
                &=
                \sum_{\alpha, \beta=0}^1
                \bigotimes_{i=1}^3&&
                \sum_{n=0}^\infty
                \left(\sqrt{2n+\alpha}\ket{2n+\alpha-1}\bra{2n+\alpha}
                + \sqrt{2n+\alpha+1}\ket{2n+\alpha+1}\bra{2n+\alpha}\right)
                \otimes P_{2\times2}(\beta) \\
                &=
                \sum_{\alpha, \beta=0}^1
                \bigotimes_{i=1}^3&&
                \Bigg[
                \sum_{n=0}^\infty
                \sqrt{2n+\alpha}\ket{2n+\alpha-1}\bra{2n+\alpha} \\
                & &&+
                \sum_{n=0}^\infty
                \sqrt{2n+\alpha+1}\ket{2n+\alpha+1}\bra{2n+\alpha}
                \Bigg]\otimes P_{2\times2}(\beta)
            \end{alignat*}
            where we understand that if $2n+\alpha-1 < 0$ then $\ket{2n+\alpha-1}=0$. We have split the summation on $n$ in two different summations. We will perform a change of variables in the first one, so that $n \rightarrow n + 1$. Note that in that case the summation index starts at -1
            \begin{alignat*}{2}
                x_1x_2x_3P &=
                \sum_{\alpha, \beta=0}^1
                \bigotimes_{i=1}^3&&
                \Bigg(
                \Bigg[
                \sum_{n=-1}^\infty
                \sqrt{2n+\alpha+2}\ket{2n+\alpha+1}\bra{2n+\alpha+2} \\
                & &&+
                \sum_{n=0}^\infty
                \sqrt{2n+\alpha+1}\ket{2n+\alpha+1}\bra{2n+\alpha}
                \Bigg]\otimes P_{2\times2}(\beta)\Bigg)
            \end{alignat*}
            Now compare both summations over $n$. They contain the same \textit{ket}, $\ket{2n + \alpha + 1}$, and have different coefficients and \textit{bras}. Those coefficients and \textit{bras} match to the result of applying the $x$ operator to the projector $P(2n + \alpha + 1)$ from the right. Therefore
            \begin{alignat*}{1}
                x_1x_2x_3P =
                \sum_{\alpha, \beta=0}^1
                \bigotimes_{i=1}^3&
                \Bigg(
                \Bigg[
                \sqrt{\alpha}\ket{\alpha-1}\bra{\alpha}
                +
                \sum_{n=0}^\infty
                P(2n +\alpha + 1)x
                \Bigg]\otimes P_{2\times2}(\beta)\Bigg)
            \end{alignat*}
            The term $\sqrt{\alpha}\ket{\alpha-1}\bra{\alpha}$ is due to one of the summations over $n$ starting at $n=-1$. That term, however, is different from zero only when $\alpha=1$. In order to regroup the term with the rest of the summations is easier to study the cases $\alpha=0$ and $\alpha=1$ separately
            \begin{alignat*}{1}
                x_1x_2x_3P &=
                \sum_{\beta=0}^1
                \bigotimes_{i=1}^3
                \Bigg(
                \Bigg[
                \sum_{n=0}^\infty
                P(2n + 1)x
                \Bigg]\otimes P_{2\times2}(\beta)\Bigg) \\
                &+\sum_{\beta=0}^1
                \bigotimes_{i=1}^3
                \Bigg(
                \Bigg[
                \ket{0}\bra{1}
                +
                \sum_{n=0}^\infty
                P(2n + 2)x
                \Bigg]\otimes P_{2\times2}(\beta)\Bigg)
            \end{alignat*}
            The second line is the one representing the case $\alpha=1$. Note that $\ket{0}\bra{1}$ is the result of applying $x$ to the projector $\ketbra{0} = P(0)$ from the right. Additionally, we can change the variable in the summation on $n$ so that $n \rightarrow n-1$ and put $P(0)x$ together with the rest of the summation
            \begin{alignat*}{1}
                x_1x_2x_3P &=
                \sum_{\beta=0}^1
                \bigotimes_{i=1}^3
                \Bigg(
                \Bigg[
                \sum_{n=0}^\infty
                P(2n + 1)x
                \Bigg]\otimes P_{2\times2}(\beta)\Bigg) \\
                &+\sum_{\beta=0}^1
                \bigotimes_{i=1}^3
                \Bigg(
                \Bigg[
                \sum_{n=0}^\infty
                P(2n)x
                \Bigg]\otimes P_{2\times2}(\beta)\Bigg)
            \end{alignat*}
            Finally, this expression can be formulated in terms of a new summation over $\alpha$
            \begin{alignat*}{1}
                x_1x_2x_3P &=
                \sum_{\alpha, \beta=0}^1
                \bigotimes_{i=1}^3
                \Bigg(
                \sum_{n=0}^\infty
                P(2n + \alpha)x
                \otimes P_{2\times2}(\beta)\Bigg) \\
                &= Px_1x_2x_3
            \end{alignat*}
            Therefore $[P, x_1x_2x_3] = 0$.

            We are missing a second step to prove that $P$ is a conserved quantity: it has to commute with the interaction hamiltonians of the qubits and modes. In order to do so, we will prove that $x_i\sigma_{x, i}P = Px_i\sigma_{x, i}$.
            \begin{alignat*}{2}
                x_i\sigma_{x, i}P =
                &\Bigg(
                \bigotimes_{j=1}^3
                \delta_{ij}x \otimes \sigma_x
                &&+
                (1-\delta_{ij}) \id \otimes \idq
                \Bigg) \times \\
                \times&\Bigg(
                \sum_{\alpha, \beta=0}^1
                \bigotimes_{j=1}^3
                \sum_{n=0}^\infty&&
                P(2n + \alpha)
                \otimes P_{2\times2}(\beta)
                \Bigg) \\
                =&
                \sum_{\alpha, \beta=0}^1
                \bigotimes_{j=1}^3
                \sum_{n=0}^\infty
                \Bigg(&&
                \delta_{ij}
                xP(2n + \alpha)\otimes \sigma_xP_{2\times2}(\beta) \\
                & &&+(1-\delta_{ij})
                P(2n+\alpha)\otimes P_{2\times2}(\beta)
                \Bigg)
            \end{alignat*}
            Now we will study the action of $x$ on $P(2n+\alpha)$
            \begin{alignat*}{2}
                x_i\sigma_{x, i}P =&
                \sum_{\alpha, \beta=0}^1
                \bigotimes_{j=1}^3
                \sum_{n=0}^\infty
                &&\Bigg(
                \delta_{ij}
                \sqrt{2n+\alpha}\ket{2n+\alpha-1}\bra{2n+\alpha}
                \otimes \sigma_xP_{2\times2}(\beta) \\
                & &&+\delta_{ij}
                \sqrt{2n+\alpha+1}\ket{2n+\alpha+1}\bra{2n+\alpha}
                \otimes \sigma_xP_{2\times2}(\beta) \\
                & &&+(1-\delta_{ij})
                P(2n+\alpha)\otimes P_{2\times2}(\beta)
                \Bigg)
            \end{alignat*}
            As it happened with $x_1x_2x_3P$, we will make a change in the variable $n$ so that $n \rightarrow n+1$ only on the first line
            \begin{alignat*}{2}
                x_i\sigma_{x, i}P =&
                \sum_{\alpha, \beta=0}^1
                \bigotimes_{j=1}^3
                &&\Bigg(
                \delta_{ij}
                \sum_{n=-1}^\infty
                \sqrt{2n+\alpha+2}\ket{2n+\alpha+1}\bra{2n+\alpha+2}
                \otimes \sigma_xP_{2\times2}(\beta) \\
                & &&+\delta_{ij}
                \sum_{n=0}^\infty
                \sqrt{2n+\alpha+1}\ket{2n+\alpha+1}\bra{2n+\alpha}
                \otimes \sigma_xP_{2\times2}(\beta) \\
                & &&+(1-\delta_{ij})
                \sum_{n=0}^\infty
                P(2n+\alpha)\otimes P_{2\times2}(\beta)
                \Bigg)
            \end{alignat*}
            The same way as before, the summation can be rewritten in terms of $P(2n + \alpha + 1)$ acting on $x$
            \begin{alignat*}{2}
                x_i\sigma_{x, i}P =&
                \sum_{\alpha, \beta=0}^1
                \bigotimes_{j=1}^3
                &&\Bigg(
                \delta_{ij}
                \sqrt{\alpha}\ket{\alpha-1}\bra{\alpha}
                \otimes \sigma_xP_{2\times2}(\beta) \\
                & &&+\delta_{ij}
                \sum_{n=0}^\infty
                P(2n+\alpha+1)x
                \otimes \sigma_xP_{2\times2}(\beta) \\
                & &&+(1-\delta_{ij})
                \sum_{n=0}^\infty
                P(2n+\alpha)\otimes P_{2\times2}(\beta)
                \Bigg)
            \end{alignat*}
            Now we will study the action of $\sigma_x$ on $P_{2\times2}(\beta)$
            \begin{align*}
                \sigma_xP_{2\times2}(\beta) = 
                \ket{\beta-1}\bra{\beta}
                +
                \ket{\beta+1}\bra{\beta}
            \end{align*}
            where we understand that if $\beta-1 < 0$ then $\ket{\beta-1}=0$ and if $\beta+1 > 1$ then $\ket{\beta+1} = 0$. Plugging this equation onto the last expression for $x_i\sigma_{x, i}P$ results in
            \begin{alignat*}{3}
                x_i\sigma_{x, i}P =&
                \sum_{\alpha, \beta=0}^1
                \bigotimes_{j=1}^3
                &&\Bigg(
                \delta_{ij}
                &&\Bigg[
                \sqrt{\alpha}\ket{\alpha-1}\bra{\alpha}
                +\sum_{n=0}^\infty
                P(2n+\alpha+1)x
                \Bigg] \otimes \\
                & && &&\otimes 
                \Big[\ket{\beta-1}\bra{\beta}
                +
                \ket{\beta+1}\bra{\beta}
                \Big]
                \\
                & &&+(1&&-\delta_{ij})
                \sum_{n=0}^\infty
                P(2n+\alpha)\otimes P_{2\times2}(\beta)
                \Bigg)
            \end{alignat*}
            By doing two different changes of variable in $\beta$ for each of the terms $\ket{\beta-1}\bra{\beta}$ and $\ket{\beta+1}\bra{\beta}$ and realizing that only one of those is non-zero for a particular value of $\beta$ one concludes that
            \begin{alignat*}{2}
                x_i\sigma_{x, i}P =&
                \sum_{\alpha, \beta=0}^1
                \bigotimes_{j=1}^3
                &&\Bigg(
                \delta_{ij}
                \sqrt{\alpha}\ket{\alpha-1}\bra{\alpha}
                \otimes P_{2\times2}(\beta)\sigma_x \\
                & &&+\delta_{ij}
                \sum_{n=0}^\infty
                P(2n+\alpha+1)x
                \otimes P_{2\times2}(\beta)\sigma_x \\
                & &&+(1-\delta_{ij})
                \sum_{n=0}^\infty
                P(2n+\alpha)\otimes P_{2\times2}(\beta)
                \Bigg)
            \end{alignat*}
            Lastly, we will study the cases $\alpha=0$ and $\alpha=1$ separately
            \begin{alignat*}{2}
                x_i\sigma_{x, i}P =&
                \sum_{\beta=0}^1
                \bigotimes_{j=1}^3
                &&\Bigg(
                \delta_{ij}
                \sum_{n=0}^\infty
                P(2n+1)x
                \otimes P_{2\times2}(\beta)\sigma_x \\
                & &&+(1-\delta_{ij})
                \sum_{n=0}^\infty
                P(2n)\otimes P_{2\times2}(\beta)
                \Bigg) \\
                +&
                \sum_{\beta=0}^1
                \bigotimes_{j=1}^3
                &&\Bigg(
                \delta_{ij}
                \ket{0}\bra{1} \otimes P_{2\times2}(\beta)\sigma_x
                +\delta_{ij}
                \sum_{n=0}^\infty
                P(2n+2)x
                \otimes P_{2\times2}(\beta)\sigma_x \\
                & &&+(1-\delta_{ij})
                \sum_{n=0}^\infty
                P(2n+1)\otimes P_{2\times2}(\beta)
                \Bigg)
            \end{alignat*}
            again, in the $\alpha=1$ case we can regroup the matrix element $\ket{0}\bra{1}$ as $P(0)x$ and combine it with the summation on $n$
            \begin{alignat*}{2}
                x_i\sigma_{x, i}P =&
                \sum_{\beta=0}^1
                \bigotimes_{j=1}^3
                &&\Bigg(
                \delta_{ij}
                \sum_{n=0}^\infty
                P(2n+1)x
                \otimes P_{2\times2}(\beta)\sigma_x \\
                & &&+(1-\delta_{ij})
                \sum_{n=0}^\infty
                P(2n)\otimes P_{2\times2}(\beta)
                \Bigg) \\
                +&
                \sum_{\beta=0}^1
                \bigotimes_{j=1}^3
                &&\Bigg(
                \delta_{ij}
                \delta_{ij}
                \sum_{n=0}^\infty
                P(2n)x
                \otimes P_{2\times2}(\beta)\sigma_x \\
                & &&+(1-\delta_{ij})
                \sum_{n=0}^\infty
                P(2n+1)\otimes P_{2\times2}(\beta)
                \Bigg)
            \end{alignat*}
            This expression can be condensed again in a summation over $\alpha$ so that
            \begin{alignat*}{2}
                x_i\sigma_{x, i}P =&
                \sum_{\alpha, \beta=0}^1
                \bigotimes_{j=1}^3
                \sum_{n=0}^\infty
                &&\Bigg(
                \delta_{ij}
                P(2n+\alpha)x
                \otimes P_{2\times2}(\beta)\sigma_x \\
                & &&+(1-\delta_{ij})
                P(2n)\otimes P_{2\times2}(\beta)
                \Bigg) \\
                =& Px_i\sigma_{x, i} && 
            \end{alignat*}
            Therefore we have proven that $[x_i\sigma_{xi}, P] = 0$.
            \end{widetext}

            Summarizing, the projector $P$ as defined in Eq. (\ref{conserved-quantity}) commutes with each of the ingredients that compose the full 3SPDC+3qubits Hamiltonian of Eq. (\ref{H}). We conclude that $P$ is a conserved quantity, and since the initial value of $\expval{P}$ for the initial state of vacuum $\ket{0g0g0g}$ is 1, it must remain one at all times. In other words, the state remains in the subspace that the projector $P$ projects on at all times, regardless of the RWA being taken or not on any interaction.
            \begin{align}
                \psi(t) = 
                \sum_{\alpha, \beta=0}^1
                \bigotimes_{i=1}^3
                \sum_{n=0}^\infty
                c_{\alpha, \beta, i, n}(t)\ket{2n+\alpha}\otimes\ket{\beta}
                \label{dynamical-subspace}
            \end{align}
                    
            We define the dynamical subspace as the subspace that contains $\psi$ at all times, that is, the image of $P$.

        \subsection{Some expectation values in the dynamical subspace}
            With a closed expression of the dynamical subspace, that is, the subspace that contains the time evolution of vacuum under the Hamiltonian (prior to any RWA), it is possible to compute some expectation values. In particular single, pairs and triplets of ladder operators, both involving the fields or the qubits.

            The expectation values of single creation operators on the modes are zero, in Eq. (\ref{dynamical-subspace}) all eigenbras of the superposition $\psi(t)$ will be orthogonal to all eigenkets of that same superposition if a photon is added to each one of them. That is, the $a^\dagger_i$ operator will produce kets with mixed parities, and there are no bras at the other side of the expectation value with mixed parities. A similar argument holds for the annihilation operators on each mode.
            \begin{align*}
                \expval{a_i} = \expval{a_i^\dagger} = 0
            \end{align*}

            The expectation values of single creation operators on the qubits are zero too, because of the same argument. 
            \begin{align*}
                \expval{\sigma^+_i} = \expval{\sigma^-_i} = 0
            \end{align*}

            The expectation values of pairs of creation or annihilation operators on modes are zero only if they act on different modes. If that is the case, the result is zero because of the same argument as before. If the operators act on the same mode, we are talking about the expectation value of the number operator, which must not be zero, as there is photon generation and that operator does not mix parities of the kets.
            \begin{align*}
                &\expval{a^\dagger_ia^\dagger_j} = 
                \expval{a_ia^\dagger_j} = 
                \expval{a^\dagger_ia_j} = 
                \expval{a_ia_j} = 0 \\
                &\text{ provided that } i \neq j
            \end{align*}

            The expectation values of pairs of ladder operators on the qubits are zero iff they act on different qubits, because of the same argument as with the modes.
            \begin{align*}
                &\expval{\sigma^+_i\sigma^+_j} =
                \expval{\sigma^+_i\sigma^-_j} =
                \expval{\sigma^-_i\sigma^+_j} =
                \expval{\sigma^-_i\sigma^-_j} = 0 \\
                &\text{ provided that } i \neq j
            \end{align*}

            The expectation values of pairs of ladder operators on one mode and on one qubit are zero only if the former acts on a mode that does not interact with the qubit the latter acts on. That is
            \begin{align*}
                &\expval{a^\dagger_i\sigma^+_j} =
                \expval{a^\dagger_i\sigma^-_j} =
                \expval{a_i\sigma^+_j} =
                \expval{a_i\sigma^-_j} = 0 \\
                &\text{ provided that } i \neq j
            \end{align*}
            The reason is the same as before, each operator will change the parity of two different pairs of modes and qubits, but will leave one pair with the previous parity.

            The expectation values of triplets of ladder operators on the modes are zero as long as they act on two modes. If that is the case, one of the ladder operators acts on one mode, and by the same argument as before, that expectation value must be zero.
            \begin{align*}
                \expval{a^\dagger_ia_ia_j} = 0 \text{ provided that } i \neq j
            \end{align*}

            With these expressions we have enough information to prove that the covariances in the fields' canonical variables and qubits' $x$ and $y$ spin variables are constant in time.

    \section{\textit{Z} spin covariances alone are not gaussian entanglement}
        \label{appendix-z-spin-covariance}
        In this section we will prove that any 3 qubit mixed state that has the same $x$ and $y$ covariaces to a separable state and only different $z$ spin covariances has no gaussian entanglement. The argument is very similar to those presented before: separable states have access to a particular range of values of the $z$ spin covariance. If general 3 qubit states have access to a bigger range of the $z$ spin covariances, then a gaussian witness paying attention to only the $z$ covariances could report entanglement. But if the separable and general ranges are the same, then no witness can tell the difference between those states with only one covariance. Then, a state that differs only in those $z$ covariances from a separable state, as is the case of the qubits state in the main text, cannot contain gaussian entanglement.

        For separable states the bound on the $z$ spin covariances is given by classical probability theory, in particular the Cauchy-Schwarz and Popoviciu's inequalities
        \begin{align*}
            \abs{\Delta^2O_iO_j}
            &\leq \sqrt{\Delta^2O_i\Delta^2O_j} \\
            &\leq \frac{1}{4}
            (\sup{O_i} - \inf{O_i})
            (\sup{O_j} - \inf{O_j})
        \end{align*}
        where $\sup{O}$ and $\inf{O}$ are bounds to the values a measurement of the observable $O$ may take. In particular for spin variables we have
        \begin{align*}
            \abs{\Delta^2 S_{z i}S_{z j}} \leq \frac{1}{4}
        \end{align*}
        The question remains whether this classical bound can be violated by some entangled state. The reader might supect that the answer is negative, as in the many years of research on entanglement, there are no Bell-like inequalities or witnesses built from covariances on only one axes. To prove that intuition consider a pure two-qubit state $\psi = \sum_{q_1=0}^1\sum_{q_2=0}^1 c_{q_1, q_2}\ket{q_1, q_2}$ and the fact that the covariances of the spin variables can be expressed in terms of the covariance of the excitation projector's covariance
        \begin{align*}
            \Delta^2S_{z i}S_{z j} &= \Delta^2 P_{e i} P_{e j} \\
            &=
            (1 - \abs{c_{10}}^2 - \abs{c_{01}}^2)\abs{c_{11}}^2
            - \abs{c_{11}}^4
            - \abs{c_{01}}^2\abs{c_{10}}^2
        \end{align*}
        where $P_{e i}$ is the projector onto the excited state of the $i$-th qubit and $c_{q_1 q_2}$ are the coefficients of a two-qubit pure state in the computational basis. It is a simple exercise to find the pure two qubit state that maximizes the covariance, which is a Bell state $\psi = \frac{1}{\sqrt{2}}\left[\ket{00}+\ket{11}\right]$ which yields a covariance $\Delta^2S_{z 1}S_{z 2}$ of $\frac{1}{4}$. Two-qubit mixed states can not violate this bound, the expectation value of a mixture is never larger than the largest of its pure components. General systems that contain two qubits cannot beat this bound either, as their expectation values will be the same as those of the reduced density matrix on the two qubits.

        Therefore, we have proven that no witness will be able to report entanglement by inspecting the $z$ covariances alone, and a state that differs from a separable state only in those covariances will not contain gaussian entanglement.
%

\end{document}